\documentstyle[aps,twocolumn,epsf]{revtex}
\begin{document}
\title{Classical Rules in Quantum Games}
\author{S.J.~van Enk and R.~Pike\\
Bell Labs, Lucent Technologies,\\
600-700 Mountain Ave,
Murray Hill NJ 07974}
\maketitle

\begin{abstract}
We consider two aspects of quantum game theory: the extent to which the quantum solution solves the original classical game, and to what extent the new solution can be obtained in a classical model.
\end{abstract}
\medskip
There have been a few dozen papers on quantum game theory (see, for instance \cite{qgames}).
All discuss versions of some classical game\cite{luce} where new rules that make explicit use of quantum mechanics lead to new solutions. 
Here we consider two aspects of quantum game theory that are important but that have been neglected so far. The first question is to what extent the quantum solution solves the underlying classical game. Ideally, when quantizing a game, one would like to leave the game unchanged and solve it using quantum operations. By analogy, Shor's algorithm\cite{shor} is quantum-mechanical but still solves the classical factoring problem. The second question is to what extent the quantum solution is really quantum mechanical in that it cannot be achieved classically. Taking again Shor's algorithm as guidance, no classical solution for the game of efficiently factoring large numbers is known, so quantum mechanics provides a truly novel solution.

Most papers on quantum game theory use a particular quantization scheme, developed in Ref.~\cite{eisert}. We argue that for that type of game the quantum solutions found are neither quantum mechanical, nor solve the classical game. To show this, we will take the example from \cite{eisert}, the quantized Prisoner's Dilemma, but the conclusions will hold for any game that is quantized in the same way. We pick Ref.~\cite{eisert} simply because it is the first paper where a general quantization procedure for games is constructed.

Consider a two-player game between Alice and Bob, where each has two moves available, denoted by $C$ (cooperate or confess) and $D$ (defect or deny). The payout matrix is
\begin{eqnarray}\label{C}
\begin{tabular}{|l||l||l|}\hline
Alice/Bob  &  C  & D \\ \hline
C & (3,3) & (0,5) \\ \hline
D & (5,0) & (1,1) \\ \hline
\end{tabular}
\end{eqnarray}
The dilemma is this: $(D,D)$ is the dominant equilibrium (neither player can profit from a {\em unilateral} change), but both players would prefer $(C,C)$.
All we need to know about the actual quantization procedure is the following: Alice and Bob indicate their choices by performing a certain operation on a qubit. The two qubits given to them are in an entangled state (otherwise, the quantum version is no different than the classical game). Alice and Bob still have choices $C$ and $D$ corresponding to the classical decisions to confess or deny. However, there are more choices available to them. In particular, there is a move, call it $Q$, that is in essence a superposition of confessing and denying.
The payout matrix including $Q$ can be constructed from Eq.~(17) of Ref.~\cite{eisert2},
\begin{eqnarray}
\label{Q}
\begin{tabular}{|l||l||l||l|}\hline
Alice/Bob  &  C  & D &Q \\ \hline
C & (3,3) & (0,5) &  (1,1) \\ \hline
D & (5,0) & (1,1) & (0,5)\\ \hline
Q & (1,1) & (5,0) & (3,3)\\ \hline
\end{tabular}
\end{eqnarray}
The entries where both Alice and Bob choose $C$ or $D$ remain the same, so that the classical version is represented as a subgame. However, the solution $(D,D)$ is clearly no longer an equilibrium point since each player can do better by {\em unilaterally} switching to $Q$.
In fact, $(Q,Q)$ is the solution of the new game defined by the payout matrix (\ref{Q}). In words, the move $Q$ has the following effect. If one party applies $Q$ and the other uses one of the two classical choices $C$ or $D$ then the payout is as if the other player had made the other classical choice $D$ or $C$. Thus the other player's classical choice can be changed by choosing the move $Q$. If both players choose $Q$ then the payout is as if both players confessed. 

Neither the above description of the move $Q$ nor the payout matrix (\ref{Q}) is quantum mechanical. So the essence of the quantized Prisoner's Dilemma is captured completely by a classical game. As noted in \cite{eisert}, there is a straightforward but inefficient classical model of the full quantum game that consists of writing down all quantum operations and wavefunctions. Since efficiency does not play a role in the Prisoner's Dilemma, this is in principle a valid model too.

The solution thus found solves (\ref{Q}), but to what extent does it solve the original game (\ref{C})? One clearly always needs new rules of some sort to find a new quantum solution to a particular given problem. 
In the case of factoring there is a genuine problem that is solved; it just so happens one may formulate it as a game. Games, however, are defined by their rules, and if you change the rules, you change the game. 
One should check whether the problem underlying the game, if there is such a problem, reasonably allows such changes of rules. In the Prisoner's Dilemma, for example, it seems to us counter to the spirit of the game to have an attorney or interrogater be helpful to the prisoners and give them an entangled state.

In general, for any two-person game defined by its payout matrix, the quantization procedure of \cite{eisert} gives rise to a new game, which may or may not have new solutions in the form of moves $Q_A,Q_B$ for Alice and Bob. That game can be represented by a classical game by simply including the new moves in a new payout matrix.  But again, it is a new game that is constructed and solved, not the original classical game.

A side issue is to what extent the quantization procedure of \cite{eisert} blurs the contrast between cooperative and noncooperative games. In noncooperative games players are not allowed to communicate, cannot enter binding agreements, and, importantly, cannot use {\em correlated} random variables.
However, by giving the players an entangled quantum state, one allows them in principle to make use of correlations present in such a state, 
violating the spirit of a noncooperative game.
Moreover, when comparing quantum and classical versions of a game one should of course not turn a noncooperative classical game into  an explicitly cooperative quantum version.  For instance, the solution to the quantum version of the three-person Prisoner's Dilemma given in \cite{benjamin} is valid only if the players enter a binding agreement to accept one of the three players to win in an {\em a priori} symmetric game. 

To end on a positive note, quantum game theory can be interesting and useful. One just has to be careful when comparing quantum games to classical games. For example, even if for a certain cooperative game one can reach the same solutions both classically and quantum-mechanically, a nontrivial question is how much communication between the players is needed to achieve these solutions\cite{qcomm}. 
Another type of quantum game worth investigating is one that exploits nonclassical correlations in entangled states, such as those that violate Bell inequalities. In the games discussed here nonlocal correlations did not play a role, in spite of the presence of entangled states, since in the end the various qubits are transported to one location where the final measurement is performed.
In more complicated and truly nonlocal games, though, those correlations may play a role.

We thank S. Benjamin, J. Eisert, P. Hayden, M. Lewenstein, and M. Wilkens  for useful discussions. They do not necessarily agree with all points raised here. We also thank Chris Fuchs, who does agree.

\end{document}